\documentclass[twocolumn, aps, prl, superscriptaddress]{revtex4-1}
\usepackage{graphicx}
\usepackage{amsfonts}
\usepackage{amsmath}
\usepackage{textcomp}
\usepackage{amssymb}
\usepackage{bm}
\usepackage{color}
\usepackage{subfigure}

\newcommand{\ket}[1]{{\left|{#1}\right\rangle}}

\newcommand{\bra}[1]{{\left\langle{#1}\right|}}
\newcommand{\ip}[2]{{\left\langle{#1|#2}\right\rangle}}

\newcommand{\e}[1]{\mathbf{e}_{#1}}

\renewcommand{\AA}{\mathbf{A}}

\newcommand{\qq}{\mathbf{q}}
\newcommand{\QQ}{\mathbf{Q}}
\newcommand{\pp}{\mathbf{p}}
\newcommand{\rr}{\mathbf{r}}
\newcommand{\kk}{\mathbf{k}}
\newcommand{\KK}{\mathbf{K}}

\begin{document}

\title{Synthetic 3D Spin-Orbit Coupling}
\author{Brandon M. Anderson}
\affiliation{Joint Quantum Institute, University of Maryland, College Park, Maryland 20742-4111, USA}
\affiliation{National Institute of Standards and Technology, Gaithersburg, Maryland 20899, USA}
\author{Gediminas Juzeli\={u}nas}
\affiliation{Institute of Theoretical Physics and Astronomy, Vilnius University, A. Goˇstauto 12, Vilnius 01108, Lithuania}
\author{Victor M. Galitski}
\affiliation{Joint Quantum Institute, University of Maryland, College Park, Maryland 20742-4111, USA}
\affiliation{Condensed Matter Theory Center, University of Maryland, College Park, Maryland 20742-4111, USA}
\author{I. B. Spielman}
\affiliation{Joint Quantum Institute, University of Maryland, College Park, Maryland 20742-4111, USA}

\begin{abstract}
We describe a method for creating a three-dimensional analogue to Rashba spin-orbit coupling in systems of ultracold atoms. This laser induced coupling uses Raman transitions to link four internal atomic states with a tetrahedral geometry, and gives rise to a Dirac point that is robust against environmental perturbations. We present an exact result showing that such a spin-orbit coupling in a fermionic system always gives rise to a molecular bound state. 
\end{abstract}

\maketitle

Recent experiments with synthetic gauge fields open the door to explore spin-orbit coupling and non-Abelian gauge fields in atomic  systems~\cite{SpielmanMag, SpielmanSO, SpielmanE, SpielmanPRL, BlochMag, GaugeReview, Jacob, SpinOrbit, Osterloh, NonAbelian}. Non-Abelian gauge fields provide rich ground state physics in bosonic systems~\cite{SOBEC,HuiStripes,HoStripes}, and enhance bound state formation in attractive fermion systems~\cite{BoundStates1,BoundStates2,BoundStates3,SoBoundStates}. However, apart from engineering cold-atom analogues to known Hamiltonians, with suitable choices of laser fields, cold atoms can be made to behave in ways that have no known analogue in solid state systems~\cite{Npod}.

In this Letter, we propose a method for synthesizing a 3D extension to Rashba spin-orbit coupling for ultracold atoms which we call Weyl spin-orbit coupling in analogy to Weyl fermions~\cite{WeylFermions}. The resulting rotationally symmetric dispersion has an infinite ground state degeneracy which covers a sphere, analogous to the Rashba Hamiltonian's circular ground state. 

We examine the consequences of such three-dimensional spin-orbit couplings: for example, the spectrum for the spherical case has a protected Dirac point that cannot be removed by any homogeneous Zeeman field. Finally, we consider the addition of interactions and exactly show that the 3D spin-orbit coupling (3DSOC) strongly enhances the binding energy of two fermions.


\begin{figure}[!t]
\centering
(a) Coupling Scheme \\
\includegraphics[width=.65 \columnwidth]{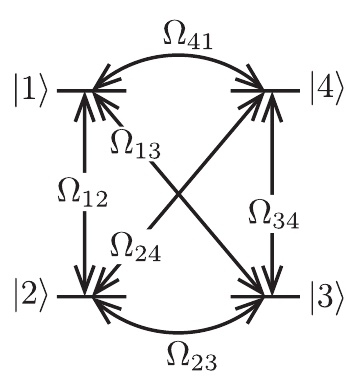}
\vspace{7 pt}
\begin{tabular}{cc}
(b) Laser Geometry  & (c) Fluxes \\
 \includegraphics[width=.45 \columnwidth]{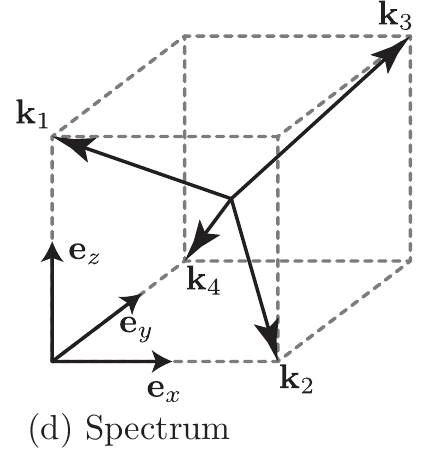} & \includegraphics[width=.45 \columnwidth]{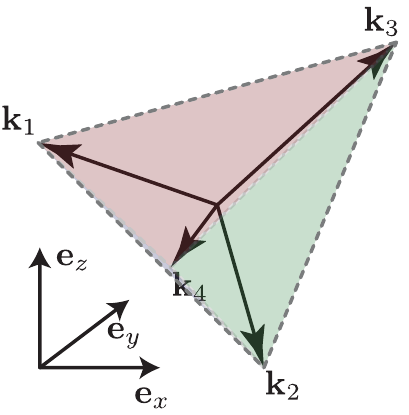}
\end{tabular}
\caption{Optical configuration for the 3DSOC. (a) Four states are coupled using two-photon optical transitions. (b) The momentum-space displacement vectors $\KK_i$ point to the four vertices of a tetrahedron. (c) The fluxes through the surface are chosen to be $\Phi_i=\pi/2$ modulo $2\pi$. }
\label{fig:tetrahedron}
\end{figure}

We produce the 3DSOC with a 4-level atom with states $\ket{1}$, $\ket{2}$, $\ket{3}$ and $\ket{4}$  optically coupled with a Hamiltonian
\begin{equation}
H_{\textrm{al}} = \sum_{jk} \Omega_{jk} \ket{j}\bra{k}
\end{equation}
with $\Omega_{jk} = \Omega^{(jk)} \exp[ i (\kk_{jk}\cdot \rr + \phi_{jk} ) ]$ as shown in Fig.~\ref{fig:tetrahedron}(a).
Here $\kk_{jk} = \KK_j - \KK_k$ is the momentum transferred by the laser, $\phi_{jk}$ is the phase of the coupling, the coupling strength $\Omega^{(jk)}$ is chosen to be $\Omega^{(1)}$ for $k=j+1$, $\Omega^{(2)}$ for $k=j+2$,  and the indexes are taken modulo 4. This coupling connects the states in a loop topology~\cite{Nlevel}, with additional next-nearest-neighbor couplings, denoted by $\Omega^{(2)}$. In the maximally symmetric case of $\Omega^{(1)} = \Omega^{(2)}$, this coupling is geometrically equivalent to a tetrahedron. We choose the momentum vectors 
\begin{eqnarray}
\KK_j & = & \kappa_\bot ( \e{x} \cos\beta_j - \e{y} \sin\beta_j ) - \kappa_\parallel (-1)^j \e{z}.
\end{eqnarray}
For the remainder of the Letter we will assume the maximally symmetric case with $k_{\|}=k_{\bot}$. The vectors $\KK_j$ point from the center to the vertices of a tetrahedron [Fig.~\ref{fig:tetrahedron}(b)]. 

The spatial dependence in the phase term can be eliminated by a state dependent boost $\ket{j} \rightarrow e^{i \KK_j \cdot \rr} \ket{j}$. In the boosted basis, the full Hamiltonian is
\begin{equation}
H = \sum_j \frac{(\pp - \KK_j)^2}{2m} \ket{j}\bra{j} + H_{\textrm{al}},
\end{equation}
with atom-laser coupling
\begin{eqnarray}
H_{al} & = & \Omega^{(1)} \sum_{j=1}^4 \left( e^{i \phi_{j,j+1}} \ket{j+1}\bra{j} +\textrm{H.C.} \right) \\
& + & \Omega^{(2)} \sum_{j=1}^2 e^{i \phi_{j,j+2}} \left( \ket{j+2}\bra{j} + \textrm{H.C.} \right)
\end{eqnarray}

The six phases $\phi_{jk}$ are not independent. Only the fluxes through the effective surfaces of the tetrahedral coupling
\begin{equation}
\Phi_i = {\sum_{k\neq i}} \phi_{k,k+1}
\end{equation}
are relevant. We choose these fluxes such that $\Phi_i = \pi/2$ modulo $2\pi$; the sum $\sum_{i=1}^4 \Phi_i=0$, so of the six $\phi_{jk}$, only three are necessary to parametrize the system. We use the additional freedom in the choice of phase to elucidate a symmetry of the problem; in what follows, we chose $\phi_{j,j+1} = \pi/4$, and $\phi_{j,j+2}=(j-1)\pi$ without loss of generality. The form of the effective Hamiltonian will be the same with any choice of phases for which the fluxes through each surface satisfy $\Phi_i = \pi / 2 \textrm{mod} 2\pi$.

We diagonalize the atom-laser Hamiltonian through a two step process. With the above choice of phases, the atom-laser Hamiltonian has a symmetry under the transformation $\ket{j} \rightarrow \ket{j+2}$. This additional symmetry allows us to dimerize states coupled by the next-nearest-neighbor couplings, $\Omega^{(2)}$ with the transformation $\ket{a_\pm} = \frac{1}{\sqrt{2}} (\ket{1} \pm \ket{3})$ and $\ket{b_\pm} = \frac{1}{\sqrt{2}} (\ket{2} \pm \ket{4})$. In the dimer basis, the atom-laser Hamiltonian is
\begin{equation}
\tilde{H}_{al} = \begin{pmatrix}
\Omega^{(2)} & \sqrt{2} \Omega^{(1)} & 0 & 0 \\
\sqrt{2} \Omega^{(1)} & - \Omega^{(2)} & 0 & 0 \\
0 & 0 & \Omega^{(2)} & -i \sqrt{2} \Omega^{(1)} \\
0 & 0 & i \sqrt{2} \Omega^{(1)} & - \Omega^{(2)}
\end{pmatrix}
\end{equation} 
with respect to the basis $\{ \ket{a_+}, \ket{b_+}, \ket{b_-}, \ket{a_-} \}$. The subsequent unitary transformation $U$ independently diagonalizes the two blocks, which have the same spectrum, $E = \pm \sqrt{2 \left[\Omega^{(1)}\right]^2 + \left[\Omega^{(2)}\right]^2}$. 
The twofold degenerate ground states comprise our pseudospin for the spin-orbit coupling. 

\begin{figure}
\centering
\begin{tabular}{cc}
(a) States &
(b) Laser Couplings \\
\includegraphics[width=.45 \columnwidth]{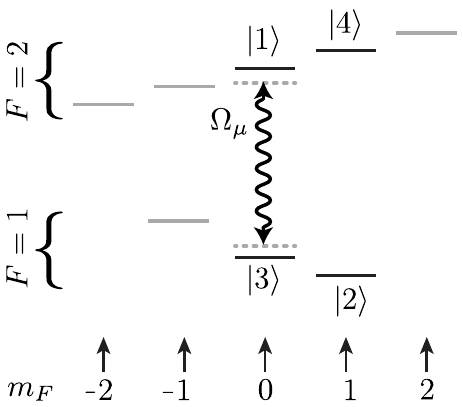} &
\includegraphics[width=.45 \columnwidth]{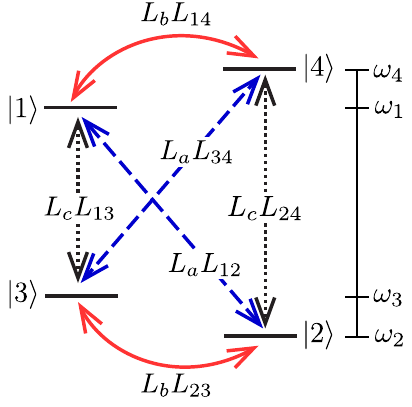}
\end{tabular} \\
(c) Laser Geometry
\includegraphics[width=.9 \columnwidth]{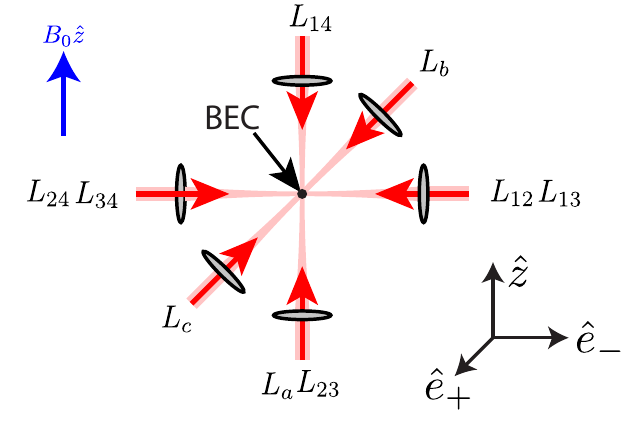}
\caption{ Four hyperfine states $\ket{F,m_F}$ of ${}^{87}{\rm Rb}$ are coupled using nine lasers. The quantization axis is set by a Zeeman field along the $\hat{z}$-axis. The couplings are produced in pairs. 
(a) The four states in the tetrahedral coupling are mapped to physical states according to $\ket{1} = \ket{2,0}$, $\ket{2} = \ket{1,+1}$, $\ket{3} = \ket{1,0}$, $\ket{1} = \ket{2,+1}$. (b) The frequencies of the three sets of lasers are given by $\{\omega_a, \omega_a + \delta_{12}, \omega_a+\delta_{34}\}$ (dashed blue), $\{\omega_b, \omega_b+\delta_{13}, \omega_b + \delta_{24}\}$ (dotted black), and $\{ \omega_c, \omega_c + \delta_{14} , \omega_c + \delta_{23} \}$ (solid red), where $\delta_{ij} = \omega_i - \omega_j$ is the frequency difference between the states $\ket{i}$ and $\ket{j}$ in the rotating frame. 
(c) The geometry of the nine laser beams. $-\kk_{12}=-\kk_{13}=\kk_{23}=\kk_{34} = k_L \hat{e}_-$, $\kk_a = \kk_{23} = -\kk_{14} = k_L \hat{z}$ and $\kk_b = - \kk_c = k_L \hat{e}_-$. The unit vectors $\hat{e}_\pm=\pm \frac{1}{\sqrt{2}}(\hat{x}\pm\hat{y})$. For a complete description of the laser parameters, see the Supplemental Material.~\cite{SuppMatScheme}  }
\label{fig:implementation}
\end{figure}

The spin-orbit coupling arises from projecting $\sum_j \qq\cdot\KK_j \ket{j}\bra{j}$ into the low energy subspace to give $\qq\cdot\AA$, where the 3D vector potential
\begin{equation}
\AA = \cos\theta \frac{\kappa_\bot}{2} (\sigma_x \mathbf{e}_x + \sigma_y \mathbf{e}_y) + \sin\theta \kappa_\parallel \sigma_z \mathbf{e}_z. \label{eq:A.sigma}
\end{equation}
nontrivially includes all three components of the Pauli matrices $ (\sigma_x,\sigma_y,\sigma_z) = \boldsymbol{\sigma} $. By changing  $\tan\theta = \Omega^{(2)}/2 \Omega^{(1)}$ and  $\kappa_\bot/\kappa_\parallel$, the vector potential can give both symmetric and asymmetric spin-orbit couplings.  The 3DSOC is fully isotropic when $\kappa_\bot / \kappa_\parallel = \Omega^{(2)} / 2 \Omega^{(1)}$, with a Hamiltonian
\begin{equation}
H_0 = \frac{\pp^2}{2m} + v \boldsymbol{\sigma} \cdot \pp. \label{eq:H0}
\end{equation}
The spin-orbit coupling is characterized by the velocity $v = \kappa_{eff} / m$, where $\kappa_{eff} = \kappa_\bot \cos(2\theta) /2$. 

The 3DSOC can be implemented in ${}^{87}{\rm Rb}$ using two-photon transitions. A possible implementation is given in Fig.~\ref{fig:implementation}. Nine laser beams with wavelength $\lambda$ are used to couple states within the $F=1$ and $F=2$ hyperfine manifolds. A Zeeman field of $B=200\textrm{mT}$ sets the quantization axis along the $\hat{z}$ direction. The remaining hyperfine transitions are isolated with a 6.8 GHz microwave field. Each pair of nonadjacent couplings is induced with three laser beams. For example, the optical couplings $\Omega_{12}$ and $\Omega_{34}$ are produced using the beams $L_{a}$, $L_{12}$ and $L_{34}$. These beams have the respective frequencies $\omega_a$, $\omega_a+\omega_1-\omega_2$ and $\omega_a+\omega_3-\omega_4$, and corresponding polarization vectors $\epsilon_a = \frac{1}{\sqrt{2}}(\hat{x}-\hat{y})$, $\epsilon_{12} = \epsilon_{34}=\hat{z}$. This pattern will ensure the system does not undergo unwanted optical transitions. The remaining four optical couplings are shown in Fig.~\ref{fig:implementation}. For complete details see the Supplemental Materials.~\cite{SuppMatScheme}

\begin{figure}[!t]
\centering
(a) Ground state manifold \\
\includegraphics[width=.7 \columnwidth]{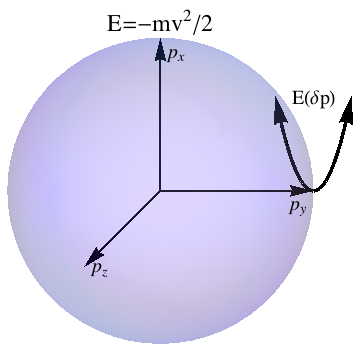}
\begin{tabular}{cc}
(b) Energy shells &  (c) Cross-section \\
\includegraphics[width=.45\columnwidth]{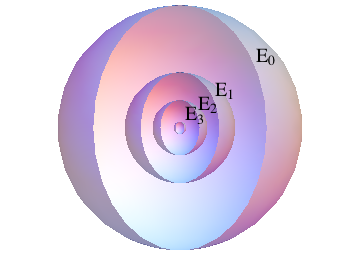} & \includegraphics[width=.45\columnwidth]{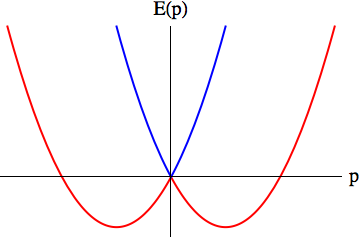}
\end{tabular}
\caption{Energy spectrum of the isotropic 3DSOC. (a) Energy spectrum of spherical ground state manifold $|\pp|= m v$ with $E=-m v^2/2$. Near the ground state manifold the dispersion $E(\delta p)$ is approximately parabolic in a small deviation, $\delta p$, from the momentum of the ground state. (b) Shells of constant energy, $E_n = - ( 1 -  0.1 n) m v^2/2 $ with $n=0,1,2,3$. These low energy shells have smaller surface area until $\pp=0$. (c) A one dimensional cut of the energy spectrum. The full spectrum is generated by rotating this spectrum along three axes. The blue(red) band corresponds to states with momentum aligned (antialigned) with spin.}
\label{fig:spectrum}
\end{figure}

The spectrum of Eq.~\ref{eq:H0} is given by
\begin{equation}
E(\pp) = \frac{\pp^2}{2m} \pm v \sqrt{p_x^2+p_y^2+p_z^2},\label{eq:E.p}
\end{equation}
and is shown in Fig~\ref{fig:spectrum}. At $\pp=0$, the system has a protected three-dimensional Dirac point. Since the spin-orbit coupling includes all spin matrices, no uniform Zeeman field can induce a splitting at the Dirac point. This is true even for anisotropic couplings, provided that the nature of the 3D spin-orbit term is preserved. 

The spectrum has a ground state manifold on a sphere $|\pp| = mv$. The dispersion is parabolic only along the radial direction, and excitations are energetically free along the polar and azimuthal directions. This suggests that a 3DSOC will behave as a quasi-one dimensional system with respect to fluctuations around the ground state. This will be manifest in the presence of a spherically symmetric trapping potential. Provided the spin-orbit energy is sufficiently strong, the low energy spectrum will be defined by a single radial quantum number, and will be degenerate in the angular quantum numbers.~\cite{DimRed}

Similarly, two free fermions with spin-orbit coupling will have a binding energy that is enhanced. It is well known that in one and two dimensions, an arbitrarily weak, attractive, potential has a bound state, whereas no bound state is guaranteed to exist in three dimensions~\cite{LandauLifshitz}. Furthermore, the binding energy in a 1D system is algebraic in the potential, while in a 2D system the binding energy is exponentially small. It was noticed~\cite{BS1,BS2}, and recently rediscovered~\cite{BoundStates1,BoundStates2,BoundStates3}, that Rashba fermions have an enhanced tendency for molecular formation. This can be understood in a manner similar to the formation of Cooper pairs, where the enhancement of the density of states near the Fermi surface reduces the effective dimension of the system from three to two. 

The situation is analogous for 3DSOC fermions, where the energetically free excitations along the polar and azimuthal direction enhances the density of states, and effectively reduces the bound state problem to one dimension. We now present the summary of an exact calculation demonstrating the binding energy of two 3D spin-orbit coupled Fermions is enhanced. For the complete calculation see the Supplemental Materials~\cite{SuppMatFermions}.  To search for bound states of two Fermions with 3D spin-orbit coupling, we solve the two-particle Schr\"{o}dinger's equation 
\begin{equation}
[H_{so}(\kk_1) \otimes \hat{1} + \hat{1} \otimes H_{so}(\kk_2)] \ket{\Psi} + V_{12} \ket{\Psi} = \Delta \ket{\Psi},
\end{equation}
where $\hat{V}_{12}$ is the two-particle interaction potential, and the tensor product implies an operator on the left operates on particle 1, and the operator on the right operates on particle 2. We assume the interaction to be purely local $s$-wave. 
This equation can be expressed in self-consistent form as
\begin{equation}
\ket{\Psi} = \hat{G} \hat{V}_{12} \ket{\Psi} \label{eq:Self-Consistent}
\end{equation}
where the Green's function is defined as
\begin{equation}
\hat{G} = [ (H_{so} - \Delta/2) \otimes \hat{1} + \hat{1} \otimes (H_{so} - \Delta/2) ]^{-1}.
\end{equation}
The ground state of a spin-orbit coupled atom will have energy $E_{so} = - mv^2 / 2$. We therefore search for solutions of (\ref{eq:Self-Consistent}) with energy $E < 2 E_{so} = - mv^2$. We define the binding energy as $\varepsilon = -mv^2 - \Delta > 0$.


The $s$-wave character of $\hat{V}_{12}$ will project the ground state into the singlet channel. Upon integration over the relative momentum coordinate $\kk = \kk_1 - \kk_2$, we can express the self-consistency equation as
\begin{equation}
v_0 \int \frac{\textrm{d}^3 \mathbf{k}}{(2 \pi)^3} \bra{\psi_s} G(\kk, \QQ) \ket{\psi_s} = 1,
\end{equation}
where $\ket{\psi_s}$ is the projection of the exact ground state into the singlet channel, $\QQ = \frac{1}{2}(\kk_1+\kk_2)$ is the center of mass momentum, and $v_0$ is the characteristic scale of the interaction. 
The bound state $\ket{\Psi_B(\QQ)}$ will appear as a solution to this equation with energy $\Delta < - 2 \left( \frac{m v^2}{2} \right)$, which is twice the ground state of of a single spin-orbit coupled fermion. We find that the binding energy of two fermions is given by 
\begin{equation}
\varepsilon = \Delta - mv^2 = - mv^2 \left( {mva_0} \right)^2
\end{equation}
at $\QQ=0$ to lowest order in $\varepsilon/ mv^2$. Thus, there exists a negative energy two-fermion bound state with energy that is algebraic in the interaction strength, consistent with the mean-field results obtained previously~\cite{BoundStates1}. This is in contrast to a system of two fermions without spin-orbit coupling where there is no bound state in three dimensions for weak attraction. The existence of this bound state results from the enhanced density of states near the 2D surface defined by $|\pp| = mv$. The additional states increase the effect of quantum fluctuations and provide an effective dimensional reduction of the system by $2$ to $D=3-2=1$. 

For nonzero center of mass momentum $\QQ$, the binding energy of the system becomes $\QQ$ dependent. For $\QQ \ll mv$ the binding energy is $\varepsilon(\QQ) = - mv^2 (mva_0)^2 - \frac{\QQ^2}{4m} + \mathcal{O}(\QQ)^4$, which is the dispersion for the center of mass of the free particle. At large momenta near $\QQ \sim mv$, the self-consistency equation no longer has a solution, and no bound state is possible. We note that the binding energy is a monotonically decreasing function of $\Delta$, and thus there is exactly one bound state for sufficiently small $\QQ$. This single bound state results from the delta function interaction. Additional bound states may be possible for other interactions, such as a square well or a $p$-wave interaction. 

In conclusion, we proposed a scheme to produce a 3DSOC using two-photon transitions to couple four atomic levels in a tetrahedral topology. In the limit of large optical power, the ground state is defined by a sphere. The origin of this coupling can be viewed as an approximation of the desired spherical momentum-space ground state manifold by a tetrahedron, whose spherical symmetry is restored in the infinite coupling limit. Such a coupling could give rise to interesting many-body systems, such as a Bose liquid~\cite{BoseLiquid}, or with the addition of suitable band gaps, Weyl fermions. Finally, we present an exact solution for the bound state energy of two fermions with a spherical 3D spin-orbit coupling. Such a bound state is found to always exist for sufficiently small center of mass momentum, and the energy of the bound state is algebraic in the interaction strength.

{\em Acknowledgements:}~ I.B.S. acknowledges the NSF through the PFC at JQI, and the ARO with funds from both the Atomtronics MURI and the DARPA OLE Program. G.J. acknowledges support by the EU FP7 project STREP NAMEQUAM and the Lithuanian Research Council project No. MIP-082/2012. V.G. was supported by US-ARO. B.A. would like to thank M. Cheng and S. Takei for helpful conversations.  

{\em Note:}~After the completion of this work, we were made aware of a similar calculation on the bound states of 3DSOC coupled fermions ~\cite{Scooped,Rashbon}, but not the possible origin of such coupling, as well as a possible implementation of 3DSOC on a lattice.~\cite{Goldman}

\bibliography{tetrahedron}

\clearpage

\section{Supplemental Materials: Fermionic Bound States}

To search for bound states of two Fermions with 3D spin-orbit coupling, we solve the two-particle Schr\"{o}dinger's equation 
\begin{equation}
[H_{so}(\kk_1) \otimes \hat{1} + \hat{1} \otimes H_{so}(\kk_2)] \ket{\Psi} + \hat{V}_{12} \ket{\Psi} = \Delta \ket{\Psi},
\end{equation}
where $\hat{V}_{12}$ is the two particle interaction potential, and the tensor product implies an operator the left operates on particle 1, and the operator on the right operates on particle 2. We assume the interaction to be purely local $s$-wave. 
This equation can be expressed in self-consistent form as
\begin{equation}
\ket{\Psi} = \hat{G} \hat{V}_{12} \ket{\Psi} \label{eq:Self-Consistent}
\end{equation}
where the Green's function is defined as
\begin{equation}
\hat{G} = [ (H_{so} - \Delta/2) \otimes \hat{1} + \hat{1} \otimes (H_{so} - \Delta/2) ]^{-1}.
\end{equation}
The ground state of a spin-orbit coupled atom will have energy $E_{so} = - mv^2 / 2$. We therefore search for solutions of (\ref{eq:Self-Consistent}) with energy $E < 2 E_{so} = - mv^2$. We define the binding energy as $\varepsilon = -mv^2 - \Delta > 0$. 

The Green's function can be calculated by applying the unitary matrix $U = U_1 \otimes U_2$, where 
\begin{equation}
U_j = \exp\left[ -i \frac{\theta_j}{2} \left( \mathbf{n}_j^* \cdot \hat{\boldsymbol{\sigma}} \right) \right]
\end{equation}
is the unitary matrix that rotates from the original spin basis to the pseudo-spin basis defined by $\ket{\alpha\beta}$, with $\alpha,\beta = \pm1$, where a particle of spin $\ket{\pm}$ has energy in the $E = \pp^2/2m \pm v |\pp|$ band. 
The vector $\mathbf{n}^*_j = ( -\sin \phi_j, \cos\phi_j, 0 )$, is perpendicular to both $\mathbf{k}_j = k_j (\cos\theta_j \cos\phi_j, \cos\theta_j\sin\theta_j, \sin\theta_j)$ and $\mathbf{e}_z$. The unitary matrix $U_j$ transforms the Green's function to
\begin{equation}
\hat{G} = \sum_{\alpha\beta} \alpha\beta d_{\alpha\beta} U \ket{\alpha\beta} \bra{\alpha\beta} U^\dagger
\end{equation}
where $d_{\alpha\beta} = \left[s + v( \alpha k_1 + \beta k_2) \right]^{-1}$.

We now assume the interaction potential is a short range, $s$-wave interaction, $\hat{V}_{12} = -v_0 \delta(\rr_1 - \rr_2) \mathcal{P}_s$, where $\mathcal{P}_s$ is a projector into the singlet state $\ket{\psi_s} = \frac{1}{2} \left( \ket{\uparrow \downarrow} - \ket{\downarrow\uparrow} \right)$.  Using (\ref{eq:Self-Consistent}), we first apply the interaction potential to the state $\ket{\Psi}$ to get $V_{12} \ip{\kk}{\Psi_B(\QQ)} = \int \frac{\textrm{d}^3\kk}{(2\pi)^3} \ip{\psi_s,\kk}{\Psi_B(\QQ)} \ket{\psi_s}$, where $\ip{\kk}{\Psi_B(\QQ)}$ is the momentum-space wavefunction of the relative coordinate $\kk = (\kk_1 - \kk_2)/2$. The center-of-mass momentum $\QQ = \kk_1 + \kk_2$ commutes with the Hamiltonain, and is thus a good quantum number which labels the bound state $\ket{\Psi_B(\QQ)}$. We can therefore express $\ket{\Psi_B(\QQ)} = N(\QQ) G(\kk,\QQ) \ket{\psi_s}$, where $N(\QQ) = \int \frac{\textrm{d}^3\kk}{(2\pi)^3} \ip{\psi_s,\kk}{\Psi_B(\QQ)}$.

We then find the full wavefunction by applying the Green's function to the state $\ket{\Psi(\QQ)}$ to get
\begin{equation}
\ket{\psi_B(\QQ)} = N(\QQ) \sum_{\alpha\beta = \pm 1} \alpha \beta d_{\alpha\beta} \chi_{\alpha\beta} \label{eq:Psi}
\end{equation}
where $\alpha,\beta = \pm 1$, and the spinors $\chi_{\alpha\beta}$ are most easily expressed in the coordinates $\kk_1,\kk_2$ as
\begin{equation}
\chi_{\alpha\beta} = 
\begin{pmatrix}
(\beta + \cos\theta_1) \sin\theta_2 e^{-i \phi_2} + (\alpha + \cos\theta_2) \sin\theta_1 e^{-i \phi_1} \\
(\alpha + \cos\theta_1)(-\beta + \cos\theta_2) -\sin\theta_1 \sin\theta_2 e^{i (\phi_2 - \phi_1)} \\
-(-\alpha + \cos\theta_1)(\beta + \cos\theta_2) + \sin\theta_1 \sin\theta_2 e^{- i (\phi_2 - \phi_1)} \\
(-\beta + \cos\theta_2) \sin\theta_1 e^{i \phi_1} + (-\alpha + \cos\theta_1) \sin\theta_2 e^{i \phi_2}
\end{pmatrix}
\end{equation}
in the basis of $( \ket{\uparrow\uparrow}, \ket{\uparrow\downarrow}, \ket{\downarrow\uparrow}, \ket{\downarrow\downarrow} )^T$.
The  $\chi_{\alpha\beta}$ form an orthogonal basis for each $\kk_1,\kk_2$, but are not normalized, since $\chi^\dagger_{\alpha^\prime\beta^\prime} \chi_{\alpha\beta} = \delta_{\alpha\alpha^\prime} \delta_{\beta\beta^\prime} c_{\alpha\beta}$, 
and $c_{\alpha\beta} = 8 \left\{ 1 + \alpha \beta \left[ \cos\theta_1 \cos\theta_2 + \cos(\phi_1 - \phi_2) \sin\theta_1 \sin\theta_2 \right] \right\}$.
In this notation, the normalization is $N(\QQ) = \left( \int \frac{\textrm{d}^2\kk}{(2\pi)^3} c_{\alpha\beta} d_{\alpha\beta}^2 \right)^{-1/2}$.

The full wavefunction allows us to calculate the binding energy as follows. We substitute (\ref{eq:Psi}) into (\ref{eq:Self-Consistent}), and then integrate over the relative momentum coordinate. Due to symmetry, the triplet components of (\ref{eq:Psi}) will vanish under integration over the relative momentum, $\int \frac{\textrm{d}^2\kk}{(2\pi)^3} \ip{\kk}{\psi_B(\QQ)} = \int \frac{\textrm{d}^2\kk}{(2\pi)^3} \mathcal{P}_s \ip{\kk}{\psi_B(\QQ)}$. Therefore, left multiplying (\ref{eq:Self-Consistent}) by $\bra{\kk} \mathcal{P}_s$, and integrating over $\kk$ allows us to express the self-consistency equation as
\begin{equation}
v_0 \int \frac{\textrm{d}^3 \mathbf{k}}{(2 \pi)^3} \bra{\psi_s} G(\kk, \QQ) \ket{\psi_s} = 1.
\end{equation}
The bound state $\ket{\Psi_B(\QQ)}$ will appear as a solution to this equation with energy $\Delta < - 2 \left( \frac{m v^2}{2} \right)$, which is twice the ground state of of a single spin-orbit coupled fermion. The integrals over the four modes $d_{\alpha\beta}$ have a linear ultraviolet divergence. To regularize the integrals we replace the interaction with a renormalized scattering length 
\begin{equation}
\frac{1}{4 \pi a_0} = \frac{1}{v_0} - \frac{1}{v_{\infty}}
\end{equation}
where $1/v_\infty = \sum_{\alpha\beta} \int \frac{d^3 \kk}{(2\pi)^3} \frac{2m}{(k + \alpha mv)^2 + (k+\beta mv)^2}$. The form of this regularizer is necessary to cure the linear divergence. A regularization scheme of $1/v_\infty \sim \int \frac{d^3 \kk}{(2\pi)^3} m/k^2$ will reduce the divergence from linear to logarithmic. Calculating the integrals using this regularization scheme above can be performed exactly, giving a binding energy 
\begin{equation}
\varepsilon = \Delta - mv^2 = - mv^2 \left( {mva_0} \right)^2
\end{equation}
at $\QQ=0$ to lowest order in $\varepsilon/ mv^2$. 

\clearpage

\section{Supplemental Materials: Implementation}

The 3DSOC can be implemented in ${}^{87}{\rm Rb}$ using two photon transitions. A possible implementation is given in Fig.~\ref{fig:implementation}. Nine laser beams with wavelength $\lambda$ are used to couple states within the $F=1$ and $F=2$ hyperfine manifolds. A Zeeman field of $B=200\textrm{mT}$ sets the quantization axis along the $\hat{z}$ direction. The remaining hyperfine transitions are isolated with a 6.8 GHz microwave field. Each pair of non-adjacent couplings is induced with three laser beams. All lasers have frequencies tuned between the $D_1$ and $D_2$ transitions. To ensure that all unwanted couplings are off resonance, we chose the base frequencies much larger than the splitting between the hyperfine levels $\omega_a, \omega_b, \omega_c \gg \delta_{ij}$. 

The six optical couplings are induced in non-adjacent pairs on the state-linkage diagram. We list the properties of the three pairs of couplings independently. In what follows we define the vectors $\hat{e}_\pm = \mp \frac{1}{\sqrt{2}} \left( \hat{x} \pm \hat{y} \right)$ and $\hbar k_L$ is the recoil momentum of the laser.

\subsection{Couplings $\Omega_{12}$ and $\Omega_{34}$}

The two couplings $\Omega_{12}$ and $\Omega_{34}$ will be induced with three lasers, denoted $L_{a}$, $L_{12}$ and $L_{34}$. The beam $L_{a}$ will be shared in the two-photon couplings. The frequencies of the three beams will be chosen such that 
\begin{eqnarray}
\omega_{12} & = & \omega_{a}+\delta_{12}\\
\omega_{34} & = & \omega_{a}+\delta_{34},
\end{eqnarray}
where $\omega_a$  is the frequency of the laser $L_a$ and $\delta_{ij}$, is the frequency splitting between the states $\ket{i}$ and $\ket{j}$. With such a configuration, the transition $\ket{1} \leftrightarrow \ket{2}$ and $\ket{3} \leftrightarrow \ket{4}$ will be on resonance, while all other dipole allowed transitions will be off resonance. The wavevectors of the lasers are given by
\begin{eqnarray}
\boldsymbol{\kappa}_{a} & = & {k_L} \hat{z}\\
\boldsymbol{\kappa}_{12} & = & -{k_L} \hat{e}_{-}\\
\boldsymbol{\kappa}_{34} & = & {k_L} \hat{e}_{-}.
\end{eqnarray}
We can check that the effective momentum transfer of the couplings are given by 
\begin{eqnarray}
\kk_{12} & = & k_L (\hat{e}_- + \hat{z}) \\
& = & \frac{k_L}{\sqrt{2}} \left( \hat{x} - \hat{y} + \sqrt{2} \hat{z} \right) \\
& = & \KK_1 - \KK_2
\end{eqnarray}
and 
\begin{eqnarray}
\kk_{34} & = & k_L (-\hat{e}_- + \hat{z}) \\
& = & \frac{k_L}{\sqrt{2}} \left( \hat{y} - \hat{x} + \sqrt{2} \hat{z} \right) \\
& = & \KK_3 - \KK_4.
\end{eqnarray}
Finally, the polarizations of the lasers will be chosen such that the beams $L_{12}$ and $L_{34}$ are linearly polarized along the $\hat{z}$ direction, while $L_a$ is $\sigma_+$ polarized. 

\begin{figure}
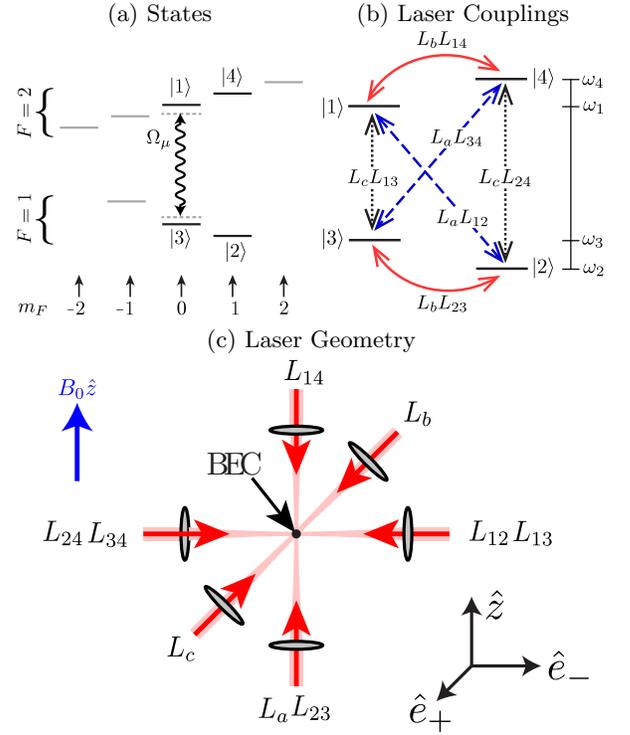

\centering
\begin{tabular}{cc}
(a) States &
(b) Laser Couplings \\
\includegraphics[width=.45 \columnwidth]{fig4_states} &
\includegraphics[width=.45 \columnwidth]{fig4_lasers}
\end{tabular} \\
(c) Laser Geometry
\includegraphics[width=.9 \columnwidth]{fig4_geometry}
\caption{Four hyperfine states $\ket{F,m_F}$ of ${}^{87}{\rm Rb}$ are coupled using nine lasers. The quantization axis is set by a Zeeman field along the $\hat{z}$-axis. The couplings are produced in pairs. 
(a) The four states in the tetrahedral coupling are mapped to physical states according to $\ket{1} = \ket{2,0}$, $\ket{2} = \ket{1,+1}$, $\ket{3} = \ket{1,0}$, $\ket{1} = \ket{2,+1}$. (b) The frequencies of the three sets of lasers are given by $\{\omega_a, \omega_a + \delta_{12}, \omega_a+\delta_{34}\}$ (dashed blue), $\{\omega_b, \omega_b+\delta_{13}, \omega_b + \delta_{24}\}$ (dotted black), and $\{ \omega_c, \omega_c + \delta_{14} , \omega_c + \delta_{23} \}$ (solid red), where $\delta_{ij} = \omega_i - \omega_j$ is the frequency difference between the states $\ket{i}$ and $\ket{j}$ in the rotating frame. 
(c) The geometry of the nine laser beams. $-\kk_{12}=-\kk_{13}=\kk_{23}=\kk_{34} = k_L \hat{e}_-$, $\kk_a = \kk_{23} = -\kk_{14} = k_L \hat{z}$ and $\kk_b = - \kk_c = k_L \hat{e}_-$. The unit vectors $\hat{e}_\pm=\pm \frac{1}{\sqrt{2}}(\hat{x}\pm\hat{y})$.}
\end{figure}

\subsection{Couplings $\Omega_{14}$ and $\Omega_{23}$}

The couplings $\Omega_{14}$ and $\Omega_{23}$ will similarly be induced with three lasers, denoted $L_{b}$, $L_{14}$ and $L_{23}$. The beam $L_{b}$ will be shared in the two-photon couplings. The frequencies of the three beams will be chosen such that 
\begin{eqnarray}
\omega_{14} & = & \omega_{b}+\delta_{14}\\
\omega_{23} & = & \omega_{b}+\delta_{23},
\end{eqnarray}
where $\omega_b$  is the frequency of the laser $L_b$. This choice of frequencies will isolate the transitions $\ket{1} \leftrightarrow \ket{4}$ and $\ket{2} \leftrightarrow \ket{3}$ in a manner similar to above couplings. The wavevectors of the lasers are given by
\begin{eqnarray}
\boldsymbol{\kappa}_{b} & = & {k_L} \hat{e}_+\\
\boldsymbol{\kappa}_{14} & = & -{k_L} \hat{z}\\
\boldsymbol{\kappa}_{23} & = & {k_L} \hat{z}.
\end{eqnarray}
We can check that the effective momentum transfer of the couplings are given by 
\begin{eqnarray}
\kk_{14} & = & k_L (\hat{e}_+ + \hat{z}) \\
& = & \frac{k_L}{\sqrt{2}} \left(\hat{x}+\hat{y}+\sqrt{2}\hat{z}\right) \\
& = & \KK_1 - \KK_4
\end{eqnarray}
and 
\begin{eqnarray}
\kk_{23} & = & k_L (\hat{e}_+ - \hat{z}) \\
& = & \frac{k_L}{\sqrt{2}} \left(\hat{x}+\hat{y}-\sqrt{2}\hat{z}\right)\\
& = & \KK_2 - \KK_3.
\end{eqnarray}
Finally, the polarizations of the lasers will be chosen such that the beams $L_{b}$ is linearly polarized along the $\hat{z}$ direction, while $L_{14}$ and $L_{23}$ are linearly polarized along the $\hat{x}$ axis. 

\subsection{Couplings $\Omega_{13}$ and $\Omega_{24} $}

The last pair of couplings $\Omega_{13}$ and $\Omega_{24}$ will similarly be induced with three lasers, denoted $L_{c}$, $L_{13}$ and $L_{24}$. The beam $L_{c}$ will be shared in the two-photon couplings. The frequencies of the three beams will be chosen such that 
\begin{eqnarray}
\omega_{13} & = & \omega_{c}+\delta_{13}\\
\omega_{24} & = & \omega_{c}+\delta_{24},
\end{eqnarray}
where $\omega_c$  is the frequency of the laser $L_c$. This choice of frequencies will isolate the transitions $\ket{1} \leftrightarrow \ket{3}$ and $\ket{2} \leftrightarrow \ket{4}$ in a manner similar to above couplings. The wavevectors of the lasers are given by
\begin{eqnarray}
\boldsymbol{\kappa}_{c} & = & - {k_L} \hat{e}_+\\
\boldsymbol{\kappa}_{13} & = & -{k_L} \hat{e}_{-}\\
\boldsymbol{\kappa}_{24} & = & {k_L} \hat{e}_{-}.
\end{eqnarray}
We can check that the effective momentum transfer of the couplings are given by 
\begin{eqnarray}
\kk_{13} & = & k_L (- \hat{e}_+ + \hat{e}_-) \\
& = & \frac{k_L}{\sqrt{2}} \left( 2 \hat{x}  \right) \\
& = & \KK_1 - \KK_3
\end{eqnarray}
and 
\begin{eqnarray}
\kk_{24} & = & k_L (- \hat{e}_+ - \hat{e}_-) \\
& = & \frac{k_L}{\sqrt{2}} \left( 2 \hat{y}  \right) \\
& = & \KK_2 - \KK_4.
\end{eqnarray}
Finally, the polarizations of the lasers will be chosen such that the beams $L_{12}$ and $L_{34}$ are linearly polarized along the $\hat{e}_+$ direction, while $L_c$ is linearly polarized along $\hat{e}_-$. 

\subsection{Amplitude and Phase}
In each pair of transitions, the three lasers provide a sufficient number of both amplitude and phase degrees of freedom to chose the the values of the couplings as desired in the main text.

\end{document}